\documentclass[12pt]{article}
\usepackage{epsfig}
\newlength{\defbaselineskip}
\newcommand{\setlinespacing}[1]%
           {\setlength{\baselineskip}{#1 \defbaselineskip}}

\begin{document}

\title{Cold nuclear matter effects and a modified form of the proximity approach}

\author{Reza Gharaei \thanks{{r.gharaei@hsu.ac.ir}}  \\
\\
{\small {\em Department of Physics, Sciences Faculty, Hakim Sabzevari University}}\\
{\small {\em P. O. Box 397, Sabzevar, Iran}}\\
}
\date{}
\maketitle

\begin{abstract}
\noindent The influence of the cold nuclear matter effects on the
Coulomb barriers and also on the fusion cross sections of 47 fusion
reactions are systematically investigated within the framework of
the proximity formalism. For this purpose, I modify the original
version of this formalism (Prox. 77) using a new analytical form of
the universal function which is formulated based on the
double-folding model with three density-dependent versions of the
M3Y-type interactions, namely DDM3Y1, CDM3Y3 and BDM3Y1. It is found
that when the Prox. 77 potential is accompanied by each of the
formulated universal functions, the agreement between the
theoretical and experimental data of the barrier height and also the
fusion cross section increase for our selected fusion systems. The
present study also provides appropriate conditions to explore
theoretically the variation effects of the NM incompressibility
constant $K$ on the calculated results caused by the Prox. 77 model.
It is shown that the accuracy of this potential model for
reproducing the experimental fusion data enhances by increasing the
strength of this constant. A discussion is also presented about the
role of various physical effects in the theoretical results of the
proximity approach.
\\
\\
PACS number(s): 25.70.Jj, 24.10.-i

\end{abstract}
%======================================================================================

\newpage
\setlinespacing{1.5}
{\noindent \bf{I. INTRODUCTION}}\\

In recent decades, the advantage of the theoretical and experimental
tools in the context of nuclear fusion reactions have provided
favorable conditions to analyze various physical aspects of such
reactions. From the theoretical point of view, it has always been
interesting to examine the influence of nuclear properties on the
complete fusion channel of two interacting nuclei. One popular
approach to evaluate role of these properties is the proximity
formalism. The original version of this formalism has been
introduced based on the "proximity force theorem" and marked as
proximity 1977 (Prox. 77) \cite{block}. As a consequence of the
literatures, one can find out that the proximity formalism needs to
be modified for achieving more accurate predictions from the
empirical/experimental data of the fusion reactions. During the
recent years, many efforts have been done to improve the theoretical
predictions of this approach
\cite{RajKu,Rajetal,OR1,OR2,Salehi13,Def1,Def2,gamma1} In those
studies, the authors tried to analyze the influence of the physical
properties such as the thermal effects of the compound nucleus
\cite{RajKu,Rajetal,OR1,OR2,Salehi13}, the deformation effects of
the target and/or projectile \cite{Def1,Def2} and the surface
tension effects of two approaching nuclei \cite{gamma1} on the
fusion process.

I can refer to the saturation property of cold nuclear matter (NM)
as an another important issue of nuclear systems. Within the
theoretical frameworks such as the double-folding (DF) model
\cite{DF1,DF2} which is accompanied by the density-dependent (DD)
M3Y interactions \cite{Khoa1}, one can explore the role of this
effect in the different fusion systems. It is well known that the
original density-independent M3Y interaction failed to predict the
correct saturation properties of a cold NM (central nucleon density
$\rho_0\simeq 0.17$ $\rm {fm}^{-3}$ and nucleon binding energy
$B(\rho_0)\simeq 16$MeV). An explicit density dependence for this
type of effective NN force reduces the strength of interactions and
also allows to consider the mentioned properties. Within the
Hartree-Fock approach \cite{Khoa2}, D. T. Khoa and coworkers
introduced eight DD versions of the M3Y interactions which are based
upon the G-matrix elements of the Paris NN potentials \cite{Ana}. As
a consequence of the literature, these generalized DD versions give
the correct saturation properties for alpha+nucleus scattering or
elastic scattering of light nuclei \cite{Khoa1,Bra}, but with
different values for corresponding NM incompressibility constants
$K$ ranging from 170 to 270 MeV. It should be noted that when the DD
interactions are employed in the DF model generate the nuclear
potentials which are too deep at small internuclear distances. This
feature is not suitable to reproduce the physical phenomena such as
the steep-falloff in the fusion cross sections at energies far below
the Coulomb barrier. To cure this deficiency, the theoretical
studies such as Refs. \cite{FH1,FH2,FH3} suggest that this model
must be modified by simulating the repulsive core effects. The
proposed procedure can be useful to justify the surface diffuseness
anomaly in the heavy-ions reactions \cite{GHZ}.

On the one side, the systematic studies such as Refs.
\cite{OR4,Dutt,DP1,DP2} indicate that the Prox. 77 potential model
overestimates the empirical data of the barrier height in different
fusion reactions. Moreover, this potential model underestimates the
measured fusion cross sections. On the other hand, it was noted
earlier that the density-dependence effects in the NN interactions
reduce the strength of such interactions. Therefore, in the present
study, we have motivated to impose indirectly these effects on the
proximity formalism and analyze role of them in the Coulomb barrier
and the fusion cross sections caused by this formalism. Finally, we
are interested to explore the importance of the effects of changing
the constant of the cold NM incompressibility on the theoretical
results of the proximity potential. To reach these aims, we
initially calculate the nuclear potential based on the DF model
which is accompanied by three versions of M3Y interactions, namely
DDM3Y1, CDM3Y3 and BDM3Y1. Then, using these calculated nuclear
potentials and also definition of the proximity potential the
behavior of the universal function $\Phi(s)$ versus the $s$
parameter is systematically studied for 47 fusion reactions. To
analyze the energy-dependence behavior of the fusion cross sections
by considering the cold NM effects in the selected fusion systems,
we employe the one-dimensional barrier penetration model (1D-BPM).

This paper is organized as follows. Section II gives the relevant
details of the theoretical models used to calculate the interaction
potential. Sec. III is devoted to the employed procedure for
parameterization of the universal function. The role of the cold NM
effects in the theoretical predictions of the Prox. 77 potential are
also studied in this section. In Sec. IV a discussion is presented
about the influence of different physical effects on the proximity
potential. The conclusions drawn from the present analysis are given
in Sec. V.
%===============================================================================================
\\
\\
\noindent{\bf {II. THEORETICAL FRAMEWORKS FOR NUCLEUS-NUCLEUS POTENTIAL}}\\

It is well recognized that the interaction potential plays a key
role in the theoretical studies of the fusion reactions. In fact, an
appropriate potential model enables us to reproduce the experimental
data of the fusion cross sections with good precision. Nowadays, the
properties of the Coulomb interactions of two colliding nuclei are
well understood. In contrast, the introduction of a comprehensive
theoretical model for evaluating various aspects of the nuclear
forces is still as a challenge. Nevertheless, several theoretical
approaches are available to calculate the nuclear part of the total
interaction potential. In the present work, the calculations of this
part are performed using two efficient models DF and proximity
potential.
%===============================================================================================
\\
\\
\noindent{\bf {A. Double-folding approach}}\\

This model is commonly used to calculate the optical potential in
the elastic scattering \cite{DF1,OP1,OP2}. In recent years, it has
also been employed to evaluate the strength of the nuclear
interactions in different fusion systems, for example see Refs.
\cite{DFF1,DFF2,DFF3}. Using this microscopic approach, the nuclear
potential between two participant nuclei can be calculated by the
sum of the strength of the NN interactions through a six-dimensional
integral as
\begin{equation} \label{1}
V_{\rm {DF}}(\mathbf{R}) = \int d\mathbf{r}_1\int
d\mathbf{r}_2\rho_1(\mathbf{r}_1)\rho_2(\mathbf{r}_2)
\upsilon_{NN}(\mathbf{r}_{12}=\mathbf{R}+\mathbf{r}_2-\mathbf{r}_1).
\end{equation}
This relation reveals that the DF integral has two main inputs; one
is the effective NN interaction $\upsilon_{NN}$ and the other is the
density distribution of the participant nuclei
$\rho_i(\mathbf{r}_i)$. In the present study, the former part is
parameterized by employing the M3Y-Paris \cite{M3YP} effective
interaction with a finite range approximation for its exchange part.
Moreover, for parametrization of the latter part we have used a
two-parameter Fermi-Dirac (2PF) distribution function
\begin{equation} \label{2}
\rho_{\rm {2PF}}(r) = \frac{\rho_0}{1+\exp{[(r-R_0)/a_0]}},
\end{equation}
Here, $R_0$ and $a_0$ are the radial and the diffuseness parameters
of nucleus, respectively.
%===============================================================================================
\\
\\
\noindent{\bf {B. Proximity approach}}\\

Proximity force theorem \cite{block} predicts that two approaching
surfaces interact with each other via the proximity force $F(s)$ at
the short distances within 2 to 3 fm. Under these conditions, one
can evaluate the nuclear proximity potential using the following
relation
\begin{equation} \label{3}
F(s)=-\bigg(\frac{\partial V_{N}^{\rm Prox.}}{\partial s}\bigg).
\end{equation}
The final form of $V_{N}^{\rm Prox.}$ can be defined as a product of
two functions. One is dependent on the shape and geometry of the
interaction system $\rm{f(shap.,geom.)}$ and the other is universal
function $\Phi(s)$,
\begin{equation} \label{21}
V^{\rm Prox.77}_{N}(r)=4\pi\gamma
b\overline{R}~\Phi(\frac{s}{b})~~\textmd{MeV},
\end{equation}
where various parts of this relation are given as follows.
\\
$\bullet$ The mean curvature radius $\overline{R}$ is
\begin{equation} \label{21}
\overline{R}=\frac{C_1C_2}{C_1+C_2},
\end{equation}
with
\begin{equation} \label{21}
C_i=R_i\bigg[1-\frac{b^2}{R_i^2}+...\bigg] ~~~~ (i=1,2).
\end{equation}
$\bullet$ The effective sharp radius $R_i$ is
\begin{equation} \label{21}
R_i=1.28A_i^{1/3}-0.76+0.8A_{i}^{-1/3} ~~ \rm{fm} ~~~~ (i=1,2).
\end{equation}
$\bullet$ The separation distance between the half-density surfaces
of the nuclei $s$ is
\begin{equation} \label{21}
s={r-C_1-C_2},
\end{equation}
$\bullet$ The surface energy coefficient $\gamma$ is
\begin{equation} \label{22}
\gamma=\gamma_0\bigg[1-k_{\rm s}\bigg(\frac{N-Z}{N+Z}\bigg)^2\bigg]
~~~ \rm{MeVfm^{-2}},
\end{equation}
where $\gamma_0$ and $k_s$ are 0.9517 MeV/$\rm {fm}^{2}$ and 1.7826,
respectively. Moreover, $N$ and $Z$ denote the neutron and proton
numbers of the compound system.
\\
$\bullet$ The universal function $\Phi(\xi=s/b)$ of the original
version of the proximity potential is
\begin{eqnarray}
\Phi(\xi) = \left\{\begin{array}{rl}
-\frac{1}{2}(\xi-2.54)^2-0.0852(\xi-2.54)^3 ~~~\rm {for} ~~\xi\leq 1.2511 \\
-3.437\rm{exp}(-\xi/0.75)~~~~~~~~~~~~~~~~~~~\rm {for} ~~ \xi\geq 1.2511\\
\end{array} \right.
\end{eqnarray}
where $b$ is of order of 1 fm.
%===============================================================================================
\\
\\
\noindent{\bf {III. CALCULATIONS AND RESULTS}}\\

\noindent{\bf {A. Parameterization of the universal function}}\\

As pointed before, we intend to investigate indirectly the cold NM
effects on the fusion barriers caused by the Prox. 77 potential
using the DD versions of the M3Y interactions. To address this goal,
I select 47 symmetric and asymmetric fusion reactions with condition
of $48\leq Z_1Z_2 \leq 2460$ for charge product of their projectile
and target. Our lightest reaction taken is that of
$^{12}$C+$^{17}$O, whereas the heaviest one is of
$^{70}$Zn+$^{208}$Pb. Moreover, it is assumed that all colliding
nuclei to be spherical in nature.

Using Eq. (4), one can obtain the following expression for
evaluating the universal function
\begin{equation} \label{4}
\Phi(\rm s)=\frac{V_N(\rm s)}{4\pi\gamma\bar{R}b}.
\end{equation}
Here, the nuclear potential $V_N(s)$ is determined based upon the DF
model which is accompanied by the DD versions of the M3Y-type
forces. Since the ability of this approach is the prediction of the
surface interactions in the regions of around the Coulomb barrier,
the redial distance between the centers of the participant nuclei
during the fusion process is restricted to $-1 \leq \rm{s} \leq 5$
range. Moreover, we employe the radial and diffuseness parameters
presented in Table I to parameterize the density distribution
functions of the reacting nuclei in the selected fusion systems. The
values of these parameters are extracted from Refs. \cite{DFF1,Vri}.

The behavior of the calculated values of the universal function
versus the parameter $s$ have been analyzed for all considered
fusion reactions in Fig. 1. The parts of (a), (b) and (c) of this
figure are devoted to the results of DF(DDM3Y1), DF(CDM3Y3) and
DF(BDM3Y1) potential models, respectively. It is found that by
increasing the values of $s$ parameter in the considered fusion
reactions and increasing the contribution of the surface
interactions in the DF potential, the universal functions calculated
by each of the mentioned versions tend to a certain constant value.
In contrast, there are fairly significant deviations in the
overlapping regions. Here, a nonlinear (fourth-order) function is
used to parameterize the behavior of $\Phi(\rm s)$ as follows
\begin{equation} \label{5}
\Phi(\rm {s})=\textit{b}_0+\textit{b}_1\rm s+\textit{b}_2\rm
s^2+\textit{b}_3\rm s^3+\textit{b}_4\rm s^4
\end{equation}
where the values of the constant coefficients $b_i$ are listed in
Table II for each of the considered potential models.

The comparison of the present pocket formula with Eq. (10)
demonstrates that the strength of the universal function becomes
more negative by imposing the DD effects. This result is clear from
Fig. 2. Moreover, since the radial behavior of the universal
function is generally similar to the nuclear potential, the
deviations of our pocket formula from Eq. (10) in the overlapping
regions can be attributed to the intrinsic properties such as the
repulsive core effects which have not been considered in the DF
formalism. The variation effects of the cold NM incompressibility
constant on the mentioned function is quite evident. In fact, by
increasing the strength of this constant from $K=176$ MeV (in DDM3Y1
version) to $K=270$ MeV (in BDM3Y1 version), the strength of
$\Phi(\rm{s})$ reduces.
%===============================================================================================
\\
\\
\noindent{\bf {B. Fusion barrier}}\\

Using the suggested form of the universal function, one can compute
the strength of the nuclear proximity potential at various
internuclear distances. Here, the modified forms of the Prox. 77
potential are labeled as "Prox.77(DDM3Y1)", "Prox.77(CDM3Y3)" and
"Prox.77(BDM3Y1)". By adding a simple form of the Coulomb potential
as $V_C(r)=\frac{Z_1Z_2e^2}{r}$ to these modified nuclear
potentials, the total interaction potentials can be calculated for
our selected fusion systems. Fig. 3 shows the theoretical values of
the barrier height $V_B^{\rm Thero.}$ as a function of the
corresponding experimental data $V_B^{\rm Exp.}$ based on the Prox.
77 model and also its modified forms. One can observe that the cold
NM effects improve the agreement between the theoretical and
experimental data of the barrier height for our selected mass range.
Furthermore, this agreement enhances by increasing the strength of
the incompressibility constant from Prox.77(DDM3Y1) model to
Prox.77(BDM3Y1) one.
%===============================================================================================
\\
\\
\noindent{\bf {C. Fusion cross section}}\\

Another right tool for investigating the complete fusion channel of
two interaction nuclei is the fusion cross section. In the present
study, 1D-BPM \cite{Bala} is employed to calculate the theoretical
values of this quantity. It is well known that in this model the
separation distance of the reacting nuclei is considered as only
degree freedom of the interaction system. Under these conditions,
the fusion cross section $\sigma_{\rm fus}$ at center-of-mass energy
$E_{\rm c.m.}$ can be defined by following relation
\begin{equation} \label{6}
\sigma_{\rm fus}(E_{\rm c.m.})=\frac{\pi\hbar^{2}}{2\mu E_{\rm
c.m.}}\sum_{l=0}^{\infty}(2\ell+1)T_{\ell}(E_{\rm c.m.}).
\end{equation}
where $\mu$ and $T_{\ell}(E_{\rm c.m.})$ are the reduced mass of the
colliding system and the transmission coefficient through the fusion
barrier, respectively. Using the WKB approximation \cite{WKB}, the
dependence of this coefficient on the energy and the angular
momentum quantities can be extracted as follows
\begin{equation} \label{7}
T_{\ell}(E_{\rm
c.m.})=\bigg[1+\textmd{exp}\bigg(2\sqrt{\frac{2\mu}{\hbar^2}}\int_{r_{1\ell}}^{r_{2\ell}}dr[V_{0}(r)+\frac{\hbar^2\ell(\ell+1)}{2\mu
r^2}-E_{\rm c.m.}]^{1/2}\bigg)\bigg]^{-1},
\end{equation}
where $r_{1\ell}$ and $r_{2\ell}$ are classical turning points for
angular momentum $\ell$ and $V_{0}(r)$ is total potential for
$\ell=0$.

The previous studies such as Refs. \cite{OR4,DP1,DP2} reveal that
the Prox. 77 potential underestimates the measured fusion cross
sections at various barrier energies. Our investigation proves that
the imposing of the cold NM effects on this potential reduces the
fusion barrier height and resulting in one can expect that the
calculated fusion cross sections enhance. Figure 4 confirms this
result for four arbitrary colliding systems $^{16}$O+$^{208}$Pb,
$^{40}$Ca+$^{40}$Ca, $^{16}$O+$^{144}$Sm, $^{16}$O+$^{72}$Ge. In
this figure the behavior of the calculated fusion cross sections
$\sigma_{\rm fus}$ (in millibarns) are plotted as a function of
$E_{\rm c.m.}$ energy (in MeV). It is shown that the Prox. 77
(DDM3Y1), Prox. 77 (CDM3Y3) and Prox. 77 (BDM3Y1) potential models
predict separately the experimental data of the mentioned quantity
with more accuracy than the Prox. 77 model. Moreover, among three
modified proximity versions, the Prox. 77 (BDM3Y1) potential has the
best consistent with the fusion data.
%=============================================================================================================
\\
\\
\noindent{\bf {IV. COMPARISON WITH THE OTHER PHYSICAL EFFECTS}}\\

The systematic studies such as Refs. \cite{Dens,GHS1,DP3} reveal
that the theoretical results caused by the proximity formalism can
be improved by taking into account various corrective effects. In
this section, the importance of those physical effects on the
theoretical predictions of the Prox. 77 model will be compared with
the cold NM effects for our selected mass range. To achieve further
understanding, we calculate the percentage difference between the
theoretical and experimental data of $V_B$ using the following
relation
\begin{equation} \label{10}
\Delta V_B (\%)=\frac{V_B^{\rm Theor.}-V_B^{\rm Exp.}}{V_B^{\rm
Exp.}}\times 100.
\end{equation}
The calculated values of $\Delta V_B (\%)$ versus the charge product
$Z_1Z_2$ are plotted in Fig. 5. We mark the results of the previous
improved proximity models as IPM1 \cite{GHS1}, IPM2 \cite{DP3}, IPM3
\cite{Dens}. As a consequence, one can see that the discrepancy
between the theoretical and experimental data of the barrier height
reduces by imposing each of the considered effects on the Prox. 77
potential. The obtained values of chi-squared
($\chi^2=\frac{1}{N}\sum_{i=1}^{N}[ V_B^{\rm Theor.}(i)-V_B^{\rm
Exp.}(i)]$) based on the considered proximity potentials are listed
in Table III.  It is clearly visible from this table that the Prox.
77 (BDM3Y1) model reproduces the experimental data with more
accuracy than the other modified versions. Indeed, the cold NM
effects have the most important role in improving the calculated
barrier heights for our fusion reactions.

It can be attractive to explore the importance of the above physical
effects on the energy-dependent behavior of the fusion cross
sections. The results of such evaluation have been shown in Fig. 6
for three arbitrary colliding systems $^{35}$Cl+$^{54}$Fe,
$^{16}$O+$^{116}$Sn and $^{24}$Mg+$^{35}$CL. As evident from this
figure, all modified forms of the proximity potential are more
successful than the Prox. 77 model to reproduce the experimental
data of $\sigma_{\rm fus}$. Moreover, this figure reveals that the
best results belongs to the proximity potential with the corrective
effects of the cold NM.
%=============================================================================================================
\\
\\
\noindent{\bf {V. SUMMARY AND CONCLUSIONS}}\\

A systematic study of the cold NM effects, including the saturation,
density and incompressibility properties, on the various
characteristics of the complete fusion channel of 47 colliding
systems is presented using the proximity formalism. We have
introduced a parameterized form of the universal function $\Phi(s)$
using the DF model with three DD versions of the M3Y-type forces.
The quality of the original version of the proximity formalism which
is modified with the parameterized form of $\Phi(s)$ is explored for
reproducing the experimental data of the barrier height and the
fusion cross section of different fusion reactions. 1D-BPM is used
to compute the theoretical values of the fusion cross section. The
most important results of the present study can be summarized in
what follows.

(i) The behavior of the universal function $\Phi(s)$ against the $s$
parameter reveals that the strength of this function reduces by
increasing the strength of the incompressibility constant $K$ in the
DD versions of the M3Y interactions.

(ii) It is shown that the imposing of the DD effects in the
proximity formalism improves the agreement between the theoretical
and experimental data of $V_B$. Moreover, this agreement enhances by
increasing the strength of the incompressibility constant $K$ for
our selected mass range. In other words, the Prox. 77 potential
generates the best result for theoretical values of the barrier
height in a hard NM.

(iii) The analysis of the energy-dependent behavior of the fusion
cross sections illustrates that the Prox.77 (BDM3Y1) model
significantly improves the calculated values of this quantity for my
considered fusion reactions. In fact, the original version of the
proximity formalism with the corrective effects of the hard nuclear
matter incompressibility reproduces the experimental data of
$\sigma_{\rm fus.}$ with more accuracy than the medium and soft
ones.

(iv) The importance of the cold NM effects in the improvement of the
theoretical predictions of the Prox. 77 model is compared with the
other physical effects such as the temperature and $\gamma$
coefficient effects. The obtained results confirm that the
corrective effects of hard NM in comparison with those proposed in
previous works \cite{Dens,GHS1,DP3} generate the more accurate
results for barrier heights and fusion cross sections caused by the
Prox. 77 model.

%=======================================================================================================
\newpage
\noindent{\bf {FIGURE CAPTIONS}}\\
\\
Fig. 1. The behavior of the universal function $\Phi(s)$ versus the
$s$ parameter based on the (a) DF(DDM3Y1) (b) DF(CDM3Y3) and (c)
DF(BDM3Y1) potential models.
\\
\\
Fig. 2. Dependence of the universal function on the variation
effects of the NM incompressibility constant $K$. A comparison has
also been performed with the analytical form suggested in Ref.
\cite{block}.
\\
\\
Fig.3. The behavior of the theoretical barrier heights $V_B^{\rm
Theor.}$ (in MeV) as a function of their corresponding experimental
data $V_B^{\rm Exp.}$ (in MeV) based on the (a) Prox. 77, (b) Prox.
77 (DDM3Y1), (c) Prox. 77 (CDM3Y3) and (d) Prox. 77 (BDM3Y1)
proximity potentials for present fusion reactions.
\\
\\
Fig. 4. Experimental fusion excitation functions for the systems (a)
$^{16}$O+$^{208}$Pb \cite{Mor}, (b) $^{40}$Ca+$^{40}$Ca \cite{HAA},
(c) $^{16}$O+$^{144}$Sm \cite{Lei} and (d) $^{16}$O+$^{72}$Ge
\cite{EFA} compared with theoretical predictions of the Prox. 77
model and its modified forms.
\\
\\
Fig. 5. The percentage differences between the theoretical and
experimental data of the barrier height as a function of charge
product $Z_1Z_2$ for our selected mass range. They are calculated
using the Prox. 77 potential and its modified forms.
\\
\\
Fig. 6. Role of various physical effects in the fusion cross
sections based on the Prox. 77 for (a) $^{35}$Cl+$^{54}$Fe
\cite{EM}, (b) $^{16}$O+$^{116}$Sn \cite{Tri} and (c)
$^{24}$Mg+$^{35}$Cl \cite{Cav} fusion reactions.
%=======================================================================================================

\newpage
Table I. The radial ($R_0$) and diffuseness ($a_0$) parameters of
various projectile and target nuclei of our selected reactions for
parameterization of their density distribution functions in DF
integral, Eq. (1). The systems are listed with respect to their
increasing $Z_1Z_2$ values.
\begin{center}
\begin{tabular}{c c c c c c}
  \hline
  \hline
  Nucleus & $Z_1Z_2$& $R_{0(P)}$ & $a_{0(P)}$ & $R_{0(T)}$&  $a_{0(T)}$\\
  \hline
  $^{12}$C+$^{17}$O & 48 & $ 2.441^{\rm a}$ & $ 0.456^{\rm a}$ & $2.661^{\rm a}$ & $0.466^{\rm a}$\\
  $^{16}$O+$^{16}$O & 64 & $2.608^{\rm a}$ & $0.465^{\rm a}$ & $2.608^{\rm a}$ & $0.465^{\rm a}$\\
  $^{16}$O+$^{40}$Ca & 160 & $2.608^{\rm a}$ & $0.465^{\rm a}$ & $3.766^{\rm a}$ & $0.544^{\rm a}$\\
  $^{26}$Mg+$^{30}$Si & 168 & $3.05^{\rm b}$ & $0.523^{\rm b}$ & $3.252^{\rm b}$ & $0.553^{\rm b}$\\
  $^{24}$Mg+$^{35}$Cl & 204 & $ 2.980^{\rm b}$ & $0.551^{\rm b}$ & $3.476^{\rm a}$ & $0.559^{\rm a}$\\
  $^{12}$C+$^{92}$Zr & 240 & $ 2.441^{\rm a}$ & $ 0.456^{\rm a}$ & $4.913^{\rm a}$ & $0.533^{\rm a}$\\
  $^{16}$O+$^{72}$Ge & 256 & $ 2.608^{\rm a}$ & $ 0.465^{\rm a}$ & $4.450^{\rm b}$ & $0.573^{\rm b}$\\
  $^{16}$O+$^{92}$Zr & 320 & $ 2.608^{\rm a}$ & $ 0.465^{\rm a}$ & $4.913^{\rm a}$ & $0.533^{\rm a}$\\
  $^{36}$S+$^{48}$Ca & 320 & $ 3.509^{\rm a}$ & $ 0.560^{\rm a}$ & $3.7369^{\rm b}$ & $0.5245^{\rm b}$\\
  $^{9}$Be+$^{208}$Pb & 328 & $ 2.218^{\rm a}$ & $ 0.458^{\rm a}$ & $6.631^{\rm a}$ & $0.505^{\rm a}$\\
  $^{19}$F+$^{93}$Nb & 369 & $ 2.590^{\rm b}$ & $ 0.564^{\rm b}$ & $4.870^{\rm b}$ & $0.573^{\rm b}$\\
  $^{12}$C+$^{152}$Sm & 372 & $ 2.441^{\rm a}$ & $ 0.456^{\rm a}$ & $5.8044^{\rm b}$ & $0.581^{\rm b}$\\
  $^{16}$O+$^{116}$Sn & 400 & $ 2.608^{\rm a}$ & $ 0.465^{\rm a}$ & $5.538^{\rm b}$ & $0.550^{\rm b}$\\
  $^{40}$Ca+$^{40}$Ca & 400 & $ 3.766^{\rm a}$ & $ 0.544^{\rm a}$ & $3.766^{\rm a}$ & $0.544^{\rm a}$\\
  $^{40}$Ca+$^{48}$Ca & 400 & $ 3.766^{\rm a}$ & $ 0.544^{\rm a}$ & $3.7369^{\rm b}$ & $0.5245^{\rm b}$\\
  $^{48}$Ca+$^{48}$Ca & 400 & $ 3.7369^{\rm b}$ & $0.5245^{\rm b}$ & $3.7369^{\rm b}$ & $0.5245^{\rm b}$\\
  $^{27}$Al+$^{70}$Ge & 416 & $ 3.070^{\rm b}$ & $0.519^{\rm b}$ & $4.440^{\rm b}$ & $0.585^{\rm b}$\\
  $^{40}$Ca+$^{48}$Ti & 440 & $ 3.766^{\rm a}$ & $ 0.544^{\rm a}$ & $3.843^{\rm b}$ & $0.588^{\rm b}$\\
  $^{35}$Cl+$^{54}$Fe & 442 & $ 3.476^{\rm a}$ & $ 0.559^{\rm a}$ & $4.075^{\rm b}$ & $0.506^{\rm b}$\\
  $^{12}$C+$^{204}$Pb & 492 & $ 2.441^{\rm a}$ & $ 0.456^{\rm a}$ & $6.588^{\rm a}$ & $0.504^{\rm a}$\\
  $^{16}$O+$^{144}$Sm & 496 & $ 2.608^{\rm a}$ & $ 0.465^{\rm a}$ & $5.719^{\rm a}$ & $0.557^{\rm a}$\\
  $^{16}$O+$^{148}$Sm & 496 & $ 2.608^{\rm a}$ & $ 0.465^{\rm a}$ & $5.771^{\rm b}$ & $0.596^{\rm b}$\\
  $^{17}$O+$^{144}$Sm & 496 & $ 2.661^{\rm a}$ & $ 0.466^{\rm a}$ & $5.719^{\rm a}$ & $0.557^{\rm a}$\\
  $^{28}$Si+$^{92}$Zr & 560 & $ 3.140^{\rm b}$ & $ 0.537^{\rm b}$ & $4.913^{\rm a}$ & $0.533^{\rm a}$\\
  $^{32}$S+$^{89}$Y & 624 & $ 3.374^{\rm a}$ & $ 0.558^{\rm a}$ & $4.760^{\rm b}$ & $0.571^{\rm b}$\\
  $^{34}$S+$^{89}$Y & 624 & $ 3.443^{\rm a}$ & $ 0.559^{\rm a}$ & $4.760^{\rm b}$ & $0.571^{\rm b}$\\
  $^{36}$S+$^{90}$Zr & 640 & $ 3.509^{\rm a}$ & $ 0.560^{\rm a}$ & $4.878^{\rm a}$ & $0.532^{\rm a}$\\
  $^{36}$S+$^{96}$Zr & 640 & $ 3.509^{\rm a}$ & $ 0.560^{\rm a}$ & $4.922^{\rm a}$ & $0.533^{\rm a}$\\
  $^{16}$O+$^{208}$Pb & 656 & $ 2.608^{\rm a}$ & $ 0.465^{\rm a}$ & $6.631^{\rm a}$ & $0.505^{\rm a}$\\
  $^{35}$Cl+$^{92}$Zr & 680 & $ 3.476^{\rm a}$ & $ 0.559^{\rm a}$ & $4.913^{\rm a}$ & $0.533^{\rm a}$\\
  $^{16}$O+$^{186}$W & 700 & $ 2.608^{\rm a}$ & $ 0.465^{\rm a}$ & $6.580^{\rm b}$ & $0.480^{\rm b}$\\
  $^{19}$F+$^{197}$Au & 711 & $ 2.590^{\rm b}$ & $ 0.564^{\rm b}$ & $6.380^{\rm b}$ & $0.535^{\rm b}$\\
  $^{16}$O+$^{238}$U & 736 & $ 2.608^{\rm a}$ & $ 0.465^{\rm a}$ & $6.8054^{\rm b}$ & $0.605^{\rm b}$\\
  \hline
  &\\
\end{tabular}
\end{center}
(a) Based on the 2PF profile extracted from \cite{DFF1}
\\
(b) Based on the 2PF profile extracted from \cite{Vri}

\newpage
Table I. (Continued.)
\begin{center}
\begin{tabular}{c c c c c c}
  \hline
  \hline
  Nucleus & $Z_1Z_2$& $R_{0(P)}$ & $a_{0(P)}$ & $R_{0(T)}$&  $a_{0(T)}$\\
  \hline
  $^{19}$F+$^{208}$Pb & 738 & $ 2.590^{\rm b}$ & $ 0.564^{\rm b}$ & $6.631^{\rm a}$ & $0.505^{\rm a}$\\
  $^{40}$Ca+$^{90}$Zr & 800 & $ 3.766^{\rm a}$ & $ 0.544^{\rm a}$ & $4.878^{\rm a}$ & $0.532^{\rm a}$\\
  $^{40}$Ca+$^{96}$Zr & 800 & $ 3.766^{\rm a}$ & $ 0.544^{\rm a}$ & $4.922^{\rm a}$ & $0.533^{\rm a}$\\
  $^{32}$S+$^{116}$Sn & 800 & $ 3.374^{\rm a}$ & $ 0.558^{\rm a}$ & $5.358^{\rm b}$ & $0.550^{\rm b}$\\
  $^{28}$Si+$^{144}$Sm & 868 & $ 3.140^{\rm b}$ & $ 0.537^{\rm b}$ & $5.719^{\rm a}$ & $0.557^{\rm a}$\\
  $^{40}$Ca+$^{124}$Sn & 1000 & $ 3.766^{\rm a}$ & $ 0.544^{\rm a}$ & $5.490^{\rm b}$ & $0.534^{\rm b}$\\
  $^{28}$Si+$^{208}$Pb & 1148 & $ 3.140^{\rm b}$ & $ 0.537^{\rm b}$ & $6.631^{\rm a}$ & $0.505^{\rm a}$\\
  $^{40}$Ar+$^{165}$Ho & 1206 & $ 3.530^{\rm b}$ & $ 0.542^{\rm b}$ & $6.180^{\rm b}$ & $0.570^{\rm b}$\\
  $^{32}$S+$^{232}$Th & 1440 & $ 3.374^{\rm a}$ & $ 0.558^{\rm a}$ & $6.7915^{\rm b}$ & $0.571^{\rm b}$\\
  $^{48}$Ti+$^{208}$Pb & 1804 & $ 3.843^{\rm b}$ & $ 0.588^{\rm b}$ & $6.631^{\rm a}$ & $0.505^{\rm a}$\\
  $^{56}$Fe+$^{208}$Pb & 2132 & $ 4.106^{\rm b}$ & $ 0.519^{\rm b}$ & $6.631^{\rm a}$ & $0.505^{\rm a}$\\
  $^{64}$Ni+$^{208}$Pb & 2296 & $ 4.212^{\rm b}$ & $ 0.578^{\rm b}$ & $6.631^{\rm a}$ & $0.505^{\rm a}$\\
  $^{70}$Zn+$^{208}$Pb & 2460 & $ 4.409^{\rm b}$ & $ 0.583^{\rm b}$ & $6.631^{\rm a}$ & $0.505^{\rm a}$\\
  \hline
  &\\
\end{tabular}
\end{center}
(a) Based on the 2PF profile extracted from \cite{DFF1}
\\
(b) Based on the 2PF profile extracted from \cite{Vri}

\newpage
Table II. The calculated values of $b_i$ constants, Eq. (12), for
each of the considered versions of the M3Y interactions.
\begin{center}
\begin{tabular}{c c c c c c}
  \hline
  \hline
  M3Y-version & $b_0$ & $b_1$ & $b_2$ & $b_3$ & $b_4$\\
  \hline
  DDM3Y1 &-4.60490 & 3.95512& -1.25831& 0.17061& -0.00801\\
  CDM3Y3 &-4.76641 & 4.18392& -1.38053& 0.19862& -0.01310\\
  BDM3Y1 &-5.76935 & 5.59630& -2.11432& 0.36185& -0.02335\\
  \hline
  &\\
\end{tabular}
\end{center}

\newpage
Table III. The $\chi^2$ values caused by imposing various physical
effects on the barrier heights calculated by the Prox. 77 for our
considered mass range.
\begin{center}
\begin{tabular}{c c c c c c}
  \hline
  \hline
  Proximity model& $\chi^2$   \\
  \hline
  Prox. 77 & 5.215    \\
  IPM1 & 3.270\\
  IPM2 & 2.878\\
  IPM3 & 2.621\\
  Prox. 77 (BDM3Y1)& 1.438\\
  \hline
  &\\
\end{tabular}
\end{center}

%===========================================================================================================================
\newpage
\begin{figure}
\begin{center}
\includegraphics{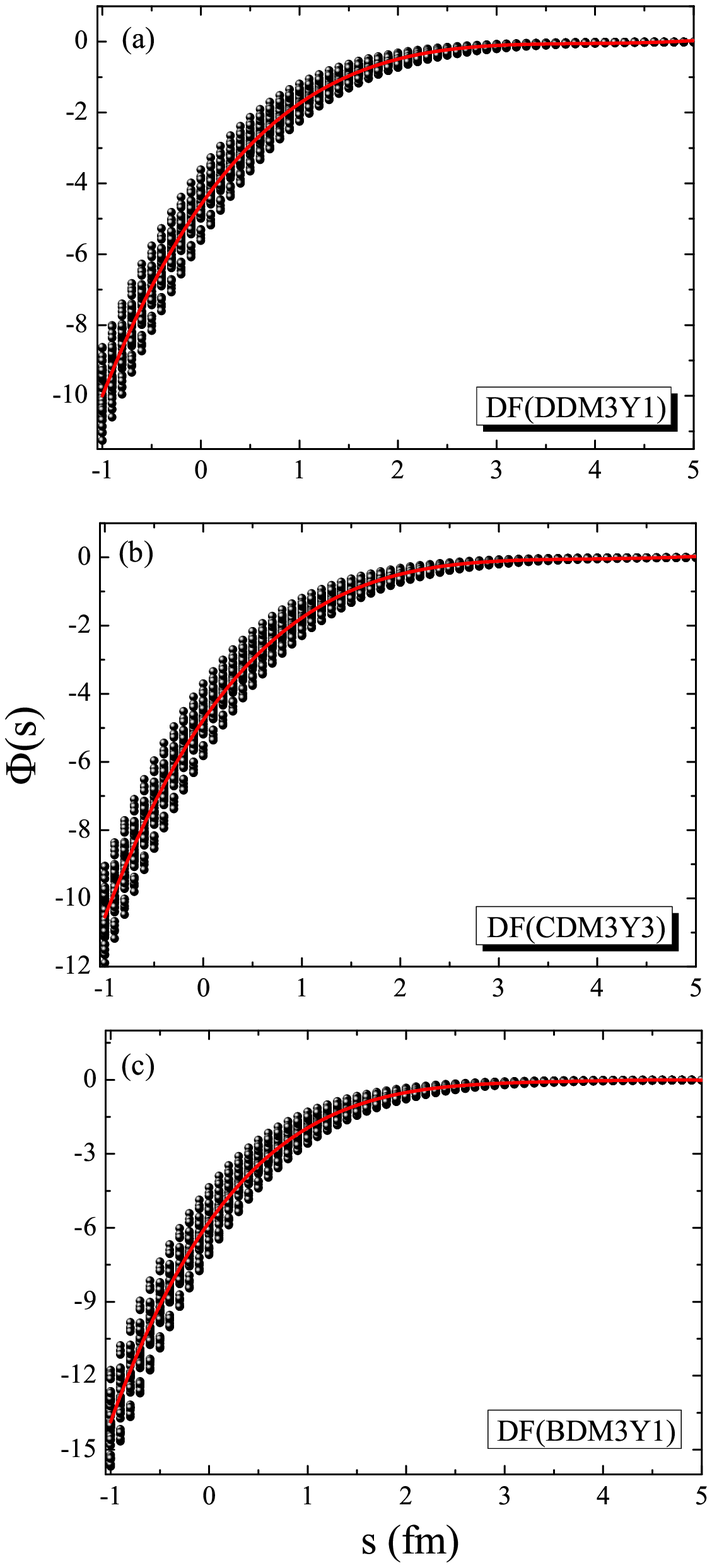}
\end{center}
\vspace{15cm} \caption{}
\end{figure}

\newpage
\begin{figure}
\begin{center}
\includegraphics{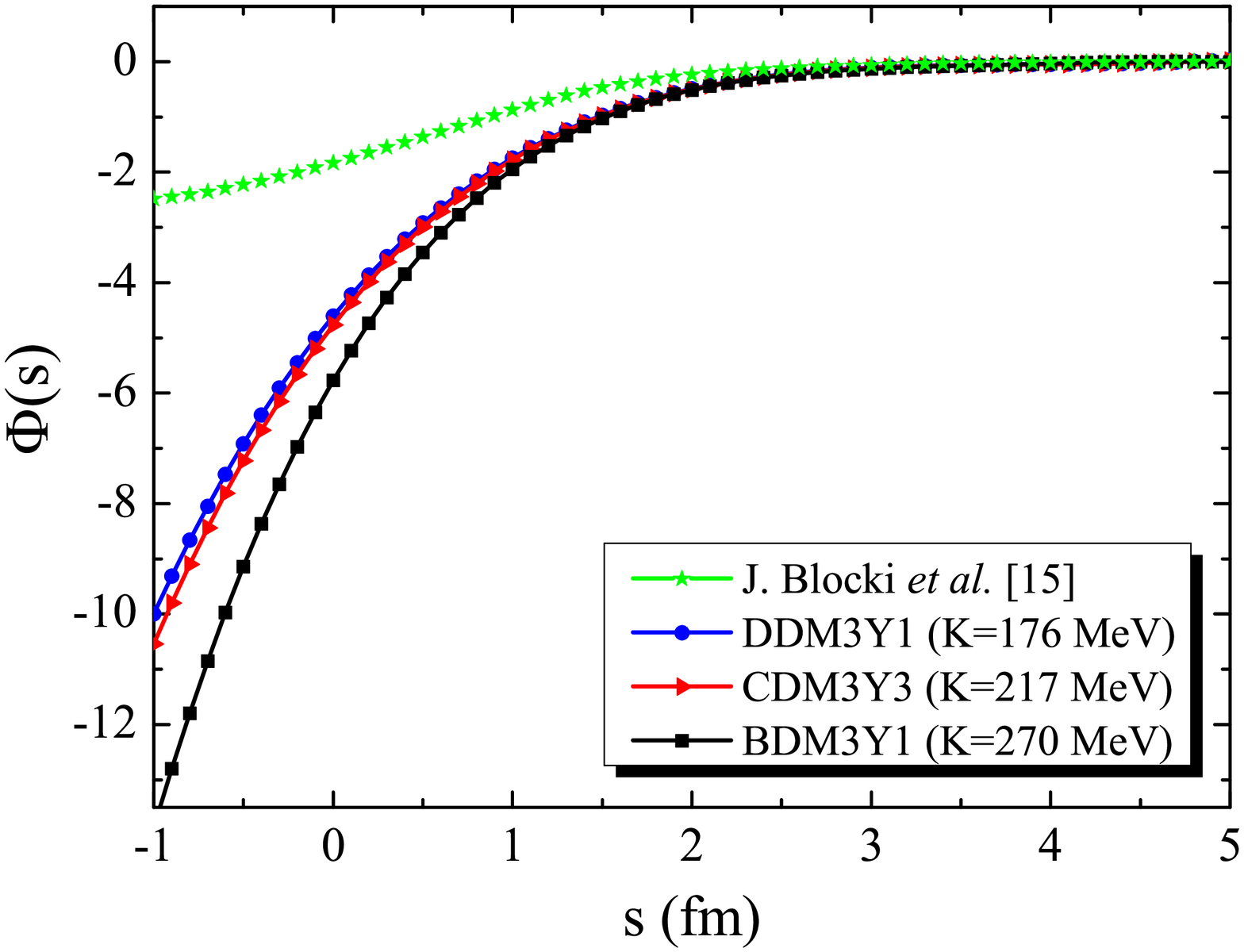}
\end{center}
\vspace{16cm} \caption{}
\end{figure}

\newpage
\begin{figure}
\begin{center}
\includegraphics{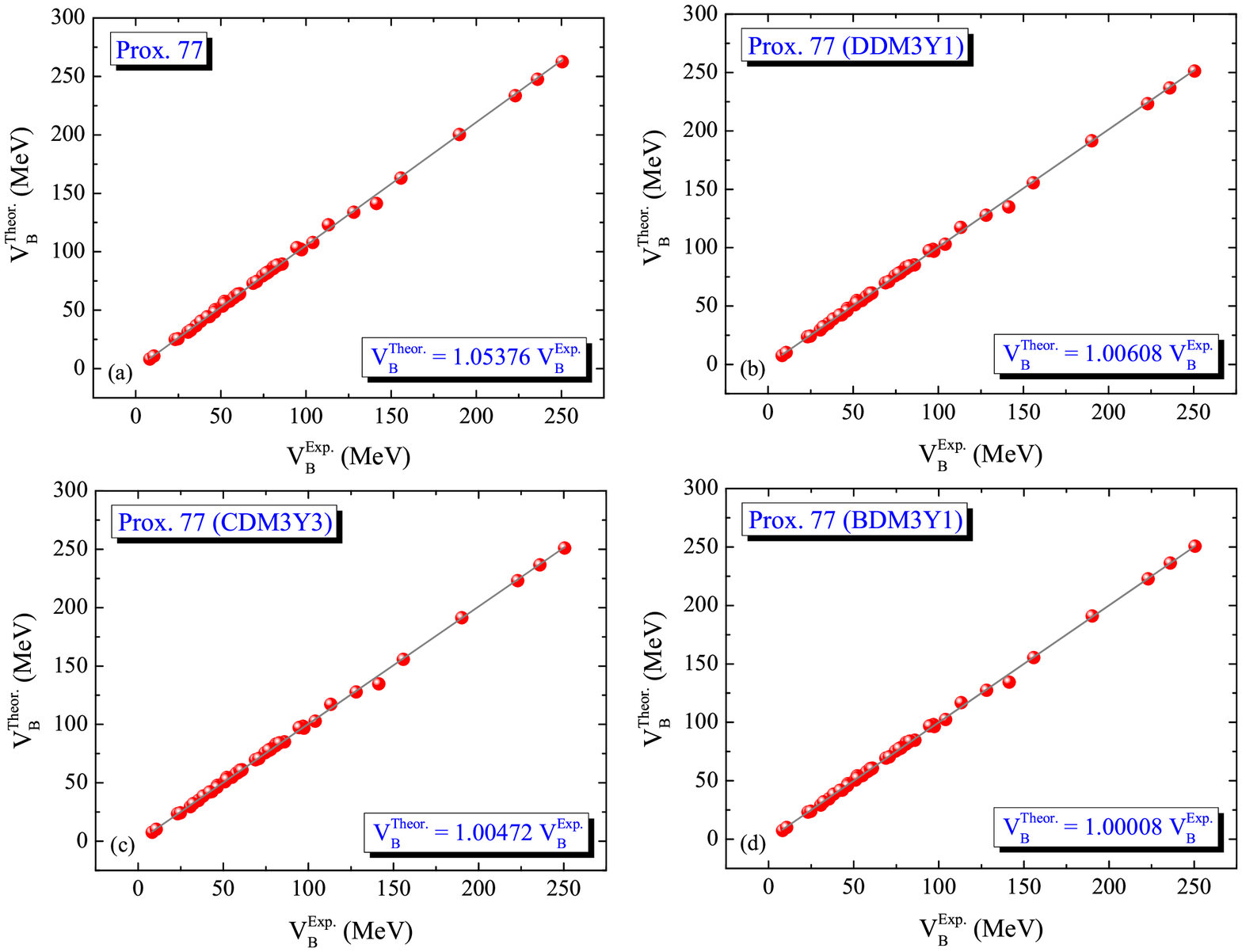}
\end{center}
\vspace{14cm} \caption{}
\end{figure}

\newpage
\begin{figure}
\begin{center}
\includegraphics{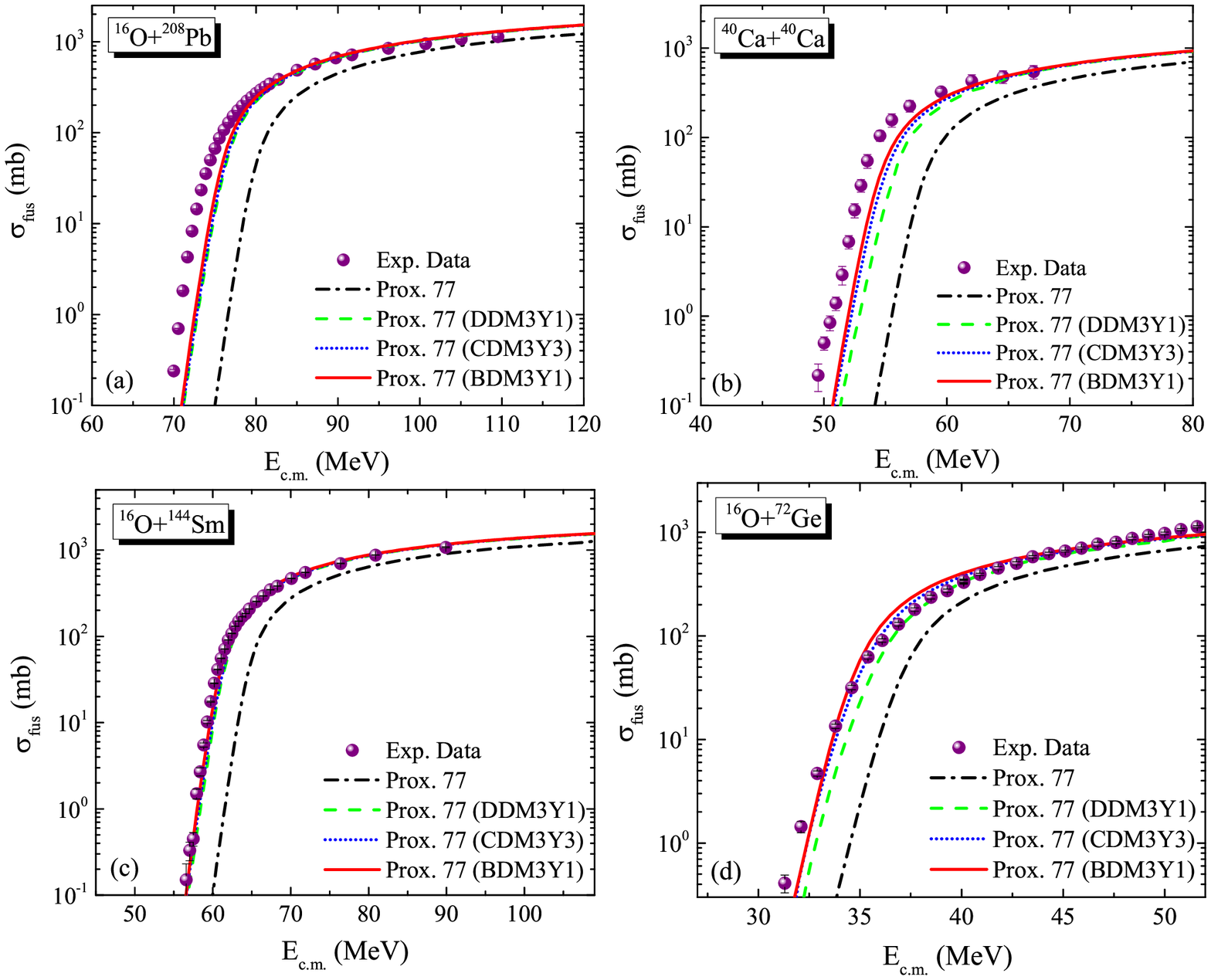}
\end{center}
\vspace{14cm} \caption{}
\end{figure}

\newpage
\begin{figure}
\begin{center}
\includegraphics{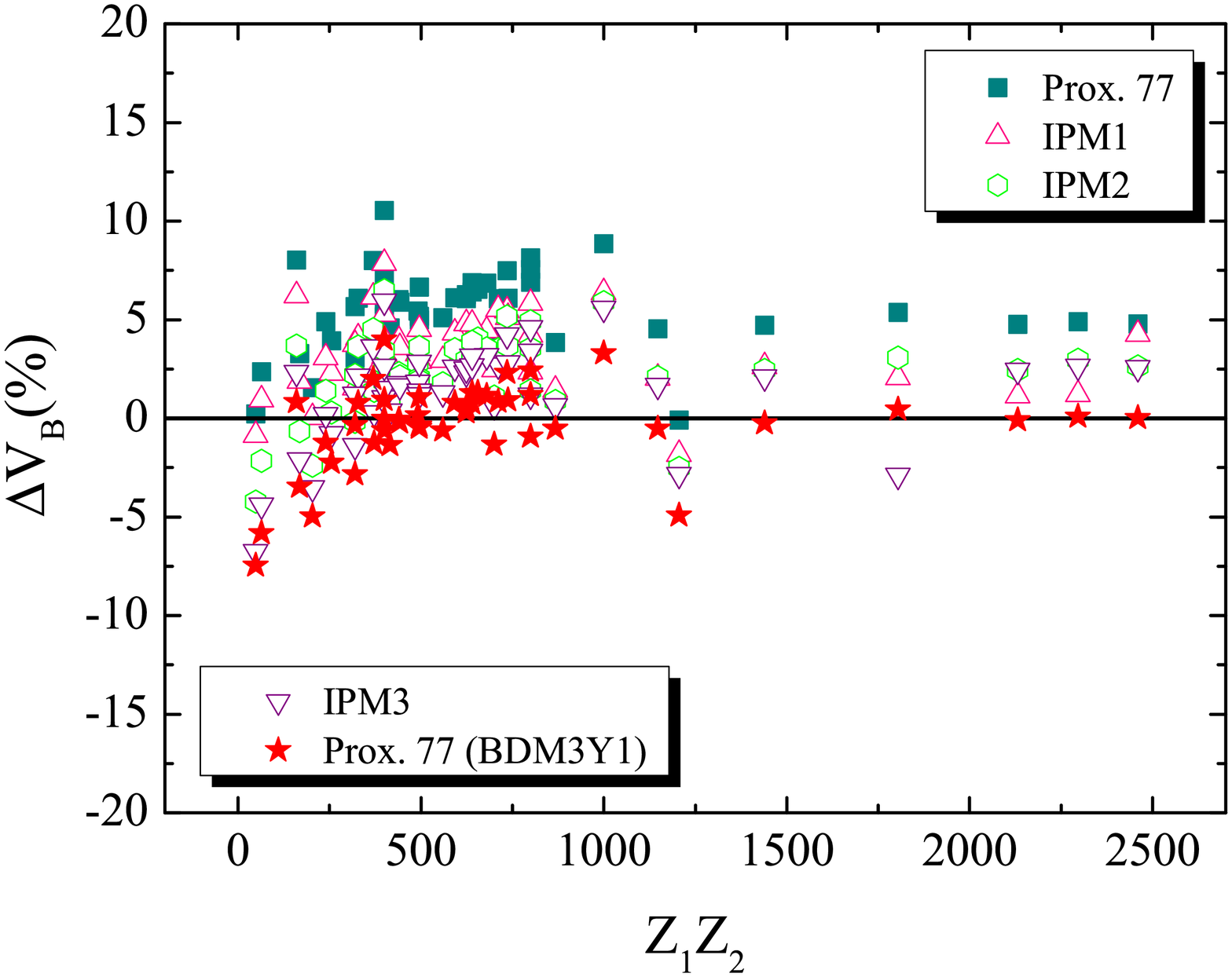}
\end{center}
\vspace{16cm} \caption{}
\end{figure}

\newpage
\begin{figure}
\begin{center}
\includegraphics{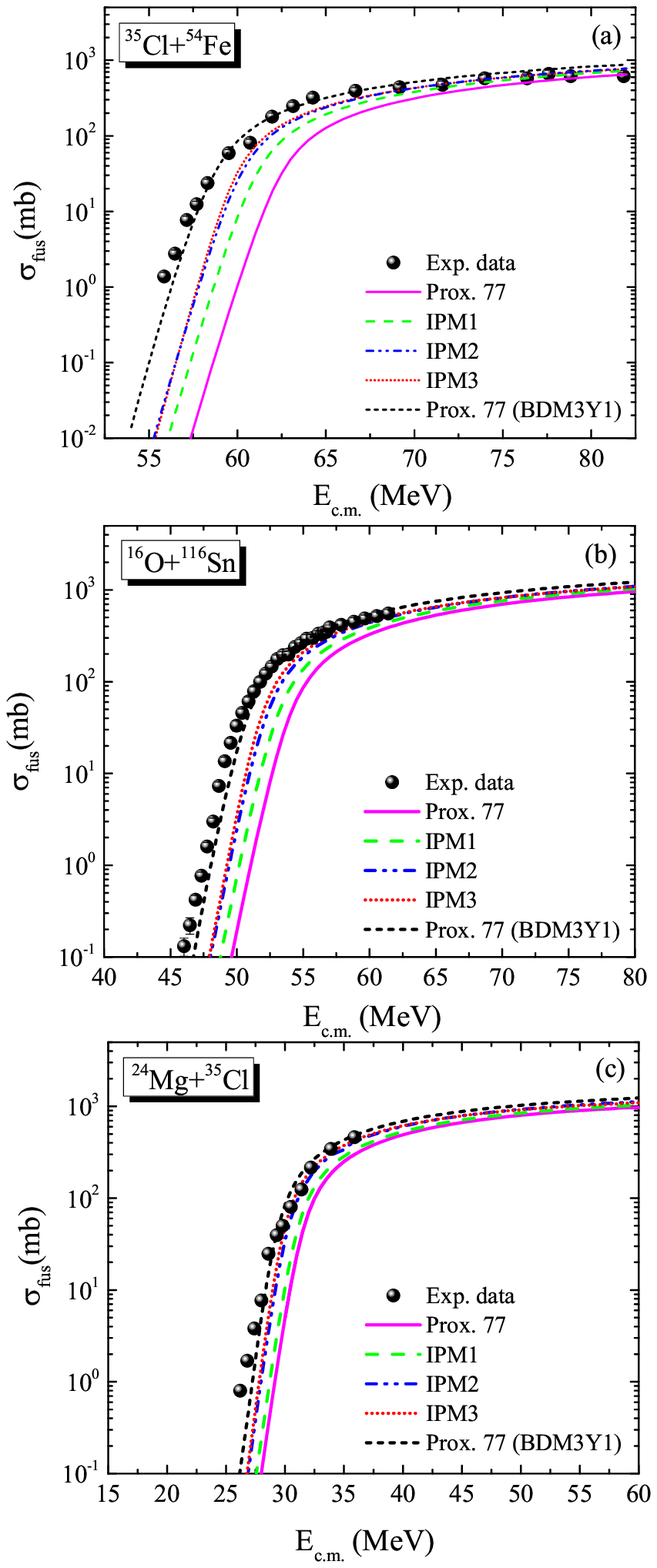}
\end{center}
\vspace{16cm} \caption{}
\end{figure}


\newpage
\begin{thebibliography}{99}

% ============================================== Refs. of Mans. ===================================================================================

\bibitem{block} J. Blocki, J. Randrup, W. J. Swiatecki and C. F. Tsang, Ann. Phys. (NY) \textbf{105}, 427 (1977).

\bibitem{RajKu} R. Kumar, Phys. Rev. C \textbf{84}, 044613 (2011).
\bibitem{Rajetal} R. Kumar, M. Bansal, S. K. Arun and R. K. Gupta, Phys. Rev. C \textbf{80}, 034618 (2009).
\bibitem{OR1} O. N. Ghodsi and R. Gharaei, Phys. Rev. C \textbf{88}, 054617 (2013).
\bibitem{OR2} O. N. Ghodsi, H. R. Moshfegh and R. Gharaei, Phys. Rev. C \textbf{88}, 034601 (2013).
\bibitem{Salehi13} M. Salehi and O. N. Ghodsi, Chin. Phys. Lett. \textbf{30}, 042502 (2013).
\bibitem{Def1} D. Jain, R. Kumar and M. K. Sharma, Nucl. Phys. A \textbf{915}, 106 (2013).
\bibitem{Def2} O. N. Ghodsi and V. Zanganeh, Phys. Rev. C \textbf{79}, 044604 (2009).
\bibitem{gamma1} I. Dutt and R. K. Puri, Phys. Rev. C \textbf{81}, 047601 (2010).

\bibitem{DF1} G. R. Satchler and W. G. Love, Phys. Rep. \textbf{55}, 183 (1979).
\bibitem{DF2} D. T. Khoa and G. R. Satchler, Nucl. Phys. A \textbf{668}, 3 (2000).
\bibitem{Khoa1} D. T. Khoa, G. R. Satchler and W. von Oertzen, Phys. Rev. C \textbf{56}, 954 (1997).
\bibitem{Khoa2} D. T. Khoa and W. von Oertzen, Phys. Lett. B \textbf{304}, 8 (1993)
\bibitem{Ana} N. Anantaraman, H. Toki and G. F. Bertsch, Nucl. Phys. A \textbf{398}, 269 (1983).
\bibitem{Bra} M. E. Brandan and G. R. Satchler, Phys. Rep. \textbf{285}, 143 (1997).

\bibitem{FH1} S. Misicu and H. Esbensen, Phys. Rev. Lett. \textbf{96}, 112701 (2006).
\bibitem{FH2} S. Misicu and H. Esbensen, Phys. Rev. C \textbf{75}, 034606 (2007).
\bibitem{FH3} E. Uegaki and Y. Abe, Prog. Theor. Phys. \textbf{90}, 615 (1993).

\bibitem{GHZ} O. N. Ghodsi and V. Zanganeh, Nucl. Phys. A \textbf{846}, 40 (2010).

\bibitem{OR4} O. N. Ghodsi and R. Gharaei, Phys. Rev. C \textbf{88}, 054617 (2013).
\bibitem{Dutt} I. Dutt and R. Bansal, Chin. Phys. Lett. \textbf{27}, 112402 (2010).
\bibitem{DP1} I. Dutt and R. K. Puri, Phys. Rev. C \textbf{81}, 044615 (2010).
\bibitem{DP2} I. Dutt and R. K. Puri, Phys. Rev. C \textbf{81}, 064609 (2010).

\bibitem{Dens} C. L. Guo, G. L. Zhang and X. Y. Le, Nucl. Phys. A \textbf{897}, 54 (2013).

\bibitem{OP1} C. W. Glover, R. I. Cutler and K. W. Kemper, Nucl. Phys. A \textbf{341}, 137 (1980).
\bibitem{OP2} M. E. Brandan and G. R. Satchler, Phys. Rep. \textbf{285}, 143 (1997).
\bibitem{DFF1} I. I. Gontchar, D. J. Hinde, M. Dasgupta and J. O. Newton, Phys. Rev. C \textbf{69}, 024610 (2004).
\bibitem{DFF2} M Rashdan, J. Phys. G: Nucl. Part. Phys. \textbf{22}, 139 (1996).
\bibitem{DFF3} W M Seif, J. Phys. G: Nucl. Part. Phys. \textbf{30}, 1231 (2004).
\bibitem{M3YP} G. Bertsch, J. Borysowicz, H. McManus and W. G. Love, Nucl. Phys. A \textbf{284}, 399 (1977).
\bibitem{Vri} H. de Vries, C. W. deJager and C. de Vries, At. Data Nucl. Data Tables \textbf{36}, 495 (1987).

\bibitem{Bala} A. B. Balantekin and N. Takigawa Reviews of Modern Physics, Vol. \textbf{70}, 77
(1998).
\bibitem{WKB} E. C. Kemble, Phys. Rev. \textbf{48}, 549 (1935).


\bibitem{GHS1} M. Salehi and O. N. Ghodsi, Chin. Phys. Lett. \textbf{30}, 042502 (2013).
\bibitem{Wan1} P. O. Biney, W. Dong and J. H. Lienhard, J. Heat Transfer, \textbf{108}, 405 (1986).
\bibitem{Wan2} I. V. J. H. Lienhard and V. J. H. Lienhard, \textit{A Heat Transfer Textbook} (Phlogiston Press), chap 9, p 465.

\bibitem{DP3} I. Dutt and R. K. Puri, Phys. Rev. C \textbf{81}, 047601 (2010).

\bibitem{Mor} C. R. Morton, A. C. Berriman, M. Dasgupta, D. J. Hinde, J. O. Newton, K. Hagino and I. J. Thompson, Phys. Rev. C \textbf{60}, 044608 (1999).
\bibitem{HAA} H. A. Aljuwair \textit{et al.}, Phys. Rev. C \textbf{30}, 1223 (1984).
\bibitem{Lei} J. R. Leigh, M. Dasgupta, D. J. Hinde \textit{et al.}, Phys. Rev. C \textbf{52}, 3151 (1995).
\bibitem{EFA} E. F. Aguilera, J. J. Kolata and R. J. Tighe, Phys. Rev. C \textbf{52}, 3103 (1995).


\bibitem{EM} E. M. Szanto, R. Liguori Neto, M. C. S. Figueira \textit{et al.}, Phys. Rev. C \textbf{41}, 2164 (1990).
\bibitem{Tri} V. Tripathi \textit{et al.}, Phys. Rev. C \textbf{65}, 014614 (2001).
\bibitem{Cav} Sl. Cavallaro, M. L. Sperduto, B. Delaunay \textit{et al.}, Nucl. Phys., A \textbf{513}, 174 (1990).

\end{thebibliography}
\end{document}